\documentclass[prl,aps,floatfix,twocolumn,superscriptaddress,preprintnumbers]{revtex4-1}

\usepackage{amsmath,amsfonts,amssymb,mathtools,cases,slashed,bm}
\usepackage{mathrsfs}
\usepackage{color}
\usepackage[dvipsnames]{xcolor}
\usepackage[colorlinks=true,citecolor=blue,linkcolor=blue,urlcolor=blue]{hyperref}
\usepackage{tabularx}
\usepackage{physics}
\usepackage[normalem]{ulem}
\usepackage{epsfig, graphicx}
\usepackage{hhline}

\usepackage{lineno}

\newcommand{\fref}[1]{Fig.~\ref{#1}}

\newcommand{\diff}[1]{\mathrm{d}#1}

\newcommand{\hel}{{^3 {\rm He}}}
\newcommand{\tri}{{^3 {\rm H}}}
\newcommand{\chidof}{{$\chi^2/N_{\rm dat}$~}}
\newcommand{\mar}{{\footnotesize MARATHON~}}
\newcommand{\Mar}{{\normalsize MARATHON~}}
\newcommand{\RR}{{\cal R}}

\begin{document}

\title{Isovector EMC effect from global QCD analysis with \Mar data}

\author{C.~Cocuzza}
\affiliation{Department of Physics, SERC, Temple University, Philadelphia, Pennsylvania 19122, USA}
\author{C.~E.~Keppel}
\affiliation{Jefferson Lab,
	     Newport News, Virginia 23606, USA}
\author{H.~Liu}
\affiliation{Department of Physics, University of Massachusetts, Amherst, Massachusetts 01003, USA}
\author{W.~Melnitchouk}
\affiliation{Jefferson Lab, Newport News, Virginia 23606, USA}
\author{A.~Metz}
\affiliation{Department of Physics, SERC, Temple University, Philadelphia, Pennsylvania 19122, USA}
\author{N.~Sato}
\affiliation{Jefferson Lab,
	     Newport News, Virginia 23606, USA}
\author{A. W. Thomas}
\affiliation{CSSM and CoEPP, Department of Physics, University of Adelaide SA 5005, Australia  \\
        \vspace*{0.2cm}
        {\bf Jefferson Lab Angular Momentum (JAM) Collaboration
        \vspace*{0.2cm} }}

\begin{abstract}
We report the results of a Monte Carlo global QCD analysis of unpolarized parton distribution functions (PDFs), including for the first time constraints from ratios of $^3$He to $^3$H structure functions recently obtained by the \mar experiment at Jefferson Lab.
Our simultaneous analysis of nucleon PDFs and nuclear effects in $A\!=\!2$ and $A\!=\!3$ nuclei reveals the first indication for an isovector nuclear EMC effect in light nuclei.
We find that while the \mar data yield relatively weak constraints on the $F_2^n/F_2^p$ neutron to proton structure function ratio and on the $d/u$ PDF ratio, they suggest an enhanced nuclear effect on the $d$-quark PDF in the bound proton, questioning the assumptions commonly made in nuclear PDF analyses.
\end{abstract}

\date{\today}
\preprint{JLAB-THY-21-3352, ADP-21-5/T1152}

\maketitle

{\it Introduction.}---\ 
The quest to unravel the 3-dimensional structure of the nucleon has recently taken on new impetus with the development of experimental programs at modern accelerator facilities at CEBAF at Jefferson Lab, RHIC at BNL and COMPASS at CERN aimed at studying processes sensitive to transverse momentum dependent  distributions and generalized parton distributions.
These complement the more traditional observables, such as from lepton-nucleon deep-inelastic scattering (DIS), that provide information on the 1-dimensional structure encoded in parton distribution functions (PDFs).
Whilst these are in contrast relatively well understood~\cite{Jimenez-Delgado:2013sma, Ethier:2020way}, even there one finds important unanswered questions.

Amongst the most notable gaps in our knowledge is the structure of valence quark PDFs carrying a large fraction $x$ ($x \to 1$) of the nucleon's light-cone momentum.
While a wealth of data has been accumulated on protons, placing significant constraints on the $u$-quark PDF, the absence of free neutron targets has meant that the $d$-quark distribution at large $x$ has remained much more elusive~\cite{Melnitchouk:1995fc, Kuhlmann:1999sf}.
The traditional method of extracting neutron structure from inclusive deuteron DIS has been shown to be handicapped by significant uncertainties in the nuclear corrections at high~$x$~\cite{Accardi:2016qay, Alekhin:2017fpf, Arrington:2011qt}.

To remedy this, several dedicated experimental efforts have been launched to map out the ratio of $d$ to $u$ PDFs at large $x$.
Amongst these are spectator proton tagging in semi-inclusive DIS from the deuteron ($D$)~\cite{Baillie:2011za, Tkachenko:2014byy}, to measure the (nearly) free neutron structure function, and $W$-boson 
production in $pp$ or $p\bar p$ collisions, which at large rapidities selects the $u$ or $d$ PDFs~\cite{Brady:2011hb}.
An alternative experiment proposed to exploit the mirror symmetry of $A=3$ nuclei to extract the neutron to proton 
structure function ratio from the ratio of $\hel$ and $\tri$ cross sections, where nuclear effects are expected to largely cancel~\cite{Afnan:2000uh, Afnan:2003vh}.
Differences between the free nucleon and nuclear structure functions were first observed by the European Muon Collaboration (EMC)~\cite{EuropeanMuon:1983wih}, which is now referred to as the ``nuclear EMC effect.''
The results from the subsequent \mar experiment that was performed at Jefferson~Lab Hall~A were recently presented~\cite{MARATHON:2021vqu}.

The experiment measured the $\hel/\tri$ ratio in the range of Bjorken-$x$ values between 0.195 and 0.825 and $Q^2$ between 2.7 and 11.9~GeV$^2$, with the $D/p$ ratio taken over a smaller $x$ range as a systematic check.
Using as an input the deuteron EMC ratio $R(D) = F_2^D/(F_2^p+F_2^n)$ or the super-ratio of the $\hel$ and $\tri$ EMC ratios,
    \mbox{$\RR = R(\hel)/R(\tri)$},
where
    $R(\hel) = F_2^{\hel}/(2 F_2^p + F_2^n)$
and
    $R(\tri) = F_2^{\tri} /(F_2^p + 2 F_2^n)$,
the neutron to proton ratio $F_2^n/F_2^p$ can be directly extracted using the measured $F_2^D/F_2^p$ or $F_2^\hel/F_2^\tri$ ratios, respectively.

In the \mar analysis~\cite{MARATHON:2021vqu}, the model calculation of Kulagin and Petti (KP)~\cite{Kulagin:2004ie} was used to extract $F_2^n/F_2^p$ from $\RR$.
Assuming that all EMC ratios for $A=2$ and $A=3$ nuclei cross unity at $x=0.31$, an overall normalization of 1.025 was applied to the $\hel/\tri$ data to match the ratio $F_2^n/F_2^p$ extracted from $F_2^D/F_2^p$ at $x=0.31$ using the KP model.
While the unity crossing is approximately established empirically from measurements of the EMC effect in heavy nuclei, $F_2^A/F_2^D$~\cite{Gomez:1993ri, Kulagin:2010gd}, it has not been demonstrated experimentally for light nuclei, with $A \leq 3$.
The \mar data are unique in their ability to provide information on isovector effects, with $n/p$ values for $\hel$ and $\tri$ ranging from $\frac12$ to 2.

%
In this Letter we use the JAM Monte Carlo global QCD framework~\cite{Sato:2019yez, Moffat:2021dji} to produce the first global analysis of the high-energy scattering data on $A=3$ nuclei, including the MARATHON results, together with data on protons and deuterons.
No prior knowledge of $\RR$ is needed, and this combined analysis allows the first {\em simultaneous} extraction of information on nucleon PDFs {\it and} nuclear effects in $A \leq 3$ nuclei.
In contrast, by assuming the KP model~\cite{Kulagin:2004ie} for the nuclear corrections, the analysis~\cite{MARATHON:2021vqu} introduces significant bias into the extracted $F_2^n/F_2^p$ ratio and underestimates the true uncertainties associated with the model dependence of the super-ratio.
In particular, while the KP model assumes that the off-shell modifications of bound protons and neutrons are equal and identical for all nuclei~\cite{Kulagin:2004ie}, our analysis allows a data-driven identification of explicit isospin dependent nuclear effects in $A=3$ systems.

{\it Theoretical framework.}---\ 
Our theoretical framework is based on the JAM iterative Monte Carlo approach to QCD global analysis~\cite{Sato:2016tuz, Moffat:2021dji}, which utilizes Bayesian inference sampling methodology that allows thorough exploration of the parameter space and robust error quantification.
Unlike attempts to extract partonic physics information from a single experiment, which invariably requires model-dependent inputs and assumptions, the virtue of a global analysis is its ability to determine the nucleon PDFs and nuclear effects simultaneously and with minimal theoretical bias.

Our analysis uses data from a variety of high-energy scattering processes, including DIS, Drell-Yan, and weak boson and jet production.
For DIS, QCD factorization theorems~\cite{Collins:1989gx} allow us to write the $F_2$ structure function as a sum of  convolutions of hard scattering functions, $C_{q,g}$, and nonperturbative quark and gluon PDFs, and higher twist (HT) power corrections,
\begin{eqnarray}
F_2^N(x,Q^2)
&=& \Big( 
    \sum_q e_q^2 \big[ C_q \otimes q_N^+ \big]
               + \big[ C_g \otimes g_N \big]
    \Big)(x,Q^2)  \notag\\
& & \times \bigg( 1 + \frac{C_N^{\rm \mbox{{\scriptsize HT}}}(x)}{Q^2} \bigg),
\label{eq.F2N}
\end{eqnarray}
where $N = p,n$, $q_N^+ \equiv q_N + \bar q_N$, and $x$ here is the Bjorken scaling variable, with $Q^2$ the four-momentum squared of the exchanged photon. 
(Note that beyond leading order in $\alpha_s$ the Bjorken variable no longer coincides with the parton momentum fraction; however, for ease of notation we will distinguish between these only when necessary.)
The hard scattering functions $C_{q,g}$ are expanded to next-to-leading order accuracy in the QCD coupling.

The coefficients of the HT terms,
    $C_N^{\rm \mbox{{\scriptsize HT}}}$,
are determined phenomenolgically from low-$Q^2$ data, and can be different for protons and neutrons.
In addition to HT corrections, at finite $Q^2$
Eq.~(\ref{eq.F2N}) includes also the effects of target mass corrections, which are
implemented within collinear factorization as described in Refs.~\cite{Aivazis:1993kh, Moffat:2019qll}.
In this analysis, we parameterize the PDFs at the input scale $Q_0^2$ using the standard form,
\begin{eqnarray}
f(x,Q_0^2) = N x^\alpha (1-x)^\beta (1 + \gamma \sqrt{x} + \eta x),
\label{eq.param}
\end{eqnarray}
as in the recent JAM19 analysis~\cite{Sato:2019yez}.

For nuclear DIS, the same factorization allows $F_2^A$ to be expressed in terms of the nuclear PDFs $q_A^+$ as in (\ref{eq.F2N}).
In the nuclear impulse approximation at $x \gg 0$ the scattering takes place incoherently from individual (off-shell) nucleons in the nucleus, and one can
write the nuclear PDF as a sum of on-shell and off-shell nucleon contributions~\cite{Melnitchouk:1993nk, Melnitchouk:1994rv, Kulagin:1994fz, Kulagin:1994cj},
\begin{eqnarray}
q_A
&=& \sum_{N} q_{N/A}
 = \sum_{N} \big[ q_{N/A}^{\rm (on)} + q_{N/A}^{\rm (off)} \big],
\label{eq.qA}
\end{eqnarray}
where we suppress the dependence upon $x$ and $Q^2$.
The notation $q_{N/A}$ refers to the PDF of a quark $q$ in a nucleus $N$, as modified within a nucleus $A$.
In the weak binding approximation (WBA)~\cite{Kulagin:1994fz, Kulagin:1994cj}, appropriate for light $A \leq 3$ nuclei, both terms in (\ref{eq.qA}) can be expressed as convolutions of nucleon smearing functions and quark distributions,
\begin{eqnarray}
q_{N/A}^{\rm (on)}
&=& f_{N/A}^{\rm (on)} \otimes q_N, \\
q_{N/A}^{\rm (off)}
&=& f_{N/A}^{\rm (off)} \otimes \delta q_{N/A},
\end{eqnarray}
where $\delta q_{N/A}$ are the off-shell smearing functions and the symbol $\otimes$ represents the convolution
    \mbox{$[f \otimes g](x)$} $\equiv \int_x^A (dy/y)\, f(x)\, g(x/y)$.
The functions $f_{N/A}^{\rm (on)}$ and $f_{N/A}^{\rm (off)}$ are on-shell and off-shell light-cone momentum distributions of nucleons $N$ in nucleus $A$, respectively~\cite{Ehlers:2014jpa, Tropiano:2018quk}, and can be computed from the nuclear wave functions or spectral functions.
Note that the integrand of the off-shell smearing function is weighted by the nucleon virtuality, $v(p^2) \equiv (p^2-M^2)/M^2 < 0$,
where $M$ is the nucleon mass, making it nearly two orders of magnitude smaller than the on-shell function.

Since the focus of the \mar experiment is on the $F_2^n/F_2^p$ (and $d/u$) ratio at large $x \gg 0$, we will restrict the discussion of the nuclear effects to the valence quark sector, which is also where the main features of the nuclear EMC effect appear.
Previous global QCD analyses with deuteron DIS data included within the WBA framework~\cite{Accardi:2016qay, Alekhin:2017fpf} found strong evidence for the presence of nucleon off-shell effects.
The off-shell corrections were implemented at the nucleon structure function level~\cite{Kulagin:2004ie, Accardi:2016qay, Alekhin:2017fpf}, with the deuteron data sensitive to one combination of the proton and neutron off-shell functions.
A later structure function analysis~\cite{Tropiano:2018quk} included $\hel/D$ ratios measured at Jefferson Lab~\cite{Seely:2009gt} and found potentially significant isospin dependence of the off-shell functions, albeit within sizeable uncertainties.
However, this nucleon level formulation with isospin dependence introduces explicit charge symmetry breaking, which one ultimately would want to test~\cite{Londergan:2009kj}.

On the other hand, by formulating the off-shell corrections at the quark level one can ensure that charge symmetry is respected.
In particular, for the deuteron the $u$ and $d$ off-shell corrections satisfy
\begin{eqnarray}
\delta u_{p/D} = \delta d_{n/D},~~~
\delta d_{p/D} = \delta u_{n/D},
\end{eqnarray}
and similarly for $\hel$ and $\tri$ nuclei,
\begin{subequations}
\begin{eqnarray}
\delta u_{p/\hel} &=& \delta d_{n/\tri},~~~
\delta d_{p/\hel}  =  \delta u_{n/\tri}, \\
\delta u_{p/\tri} &=& \delta d_{n/\hel},~~~
\delta d_{p/\tri}  =  \delta u_{n/\hel},
\end{eqnarray}
\end{subequations}
so that the 12 off-shell functions (for $u$ and $d$ quarks in $p$ and $n$ in $D$, $\hel$ and $\tri$) reduce to 6.
If we further assume that the spectator system to the DIS from $D$ and $^3$H respects isospin symmetry, then
the $u$ and $d$ off-shell functions in the deuteron and $\tri$ can be related by
\begin{subequations}
\begin{eqnarray}
\delta u_{p/D}
&=& \delta u_{p/\tri} \equiv \delta u,   \\
\delta d_{p/D}
&=& \delta d_{p/\tri} \equiv \delta d.
\end{eqnarray}
\end{subequations}
Because for $\delta u$ the product of the third component of isospin of the struck quark and of the spectator nucleon(s) is negative, while for $\delta d$ this is positive, an isovector nuclear correction would lead to changes of opposite sign between them~\cite{Mineo:2003vc, Cloet:2009qs}. 
For the off-shell corrections in the proton in $\hel$ we expect the isovector effects to approximately cancel, and take
\begin{eqnarray}
\delta u_{p/\hel}
&\approx& 2 \delta d_{p/\hel} = \frac12 \big( \delta u + 2\, \delta d \big).
\end{eqnarray}

To preserve the number of valence quarks in the bound nucleons
in nuclei, the off-shell functions must satisfy
\begin{eqnarray}
\int_0^1 \diff x~\delta u(x) = \int_0^1 \diff x~\delta d(x) = 0.
\end{eqnarray}
Note that because the off-shell functions $\delta q$ are convoluted with the off-shell smearing functions, $f_{N/A}^{\rm (off)}$, the fact that the nucleon virtuality $v(p^2)$ averages to a number roughly twice as large in magnitude in $A=3$ as $A=2$ accounts for the expected increase in off-shell effects in the former.
For the Paris~\cite{Lacombe:1981eg} deuteron wave function and the KPSV~\cite{Kievsky:1996gz} $^3$He spectral function, for example, we find
\begin{eqnarray}
\langle f_{p/D}^{\rm (off)} \rangle \approx -4.3\%,\ \
%
%
\langle f_{p/{^3}{\rm He}}^{\rm (off)} \rangle \approx -6.8\%,\ \
%
%
\langle f_{p/{^3}{\rm H}}^{\rm (off)} \rangle \approx -9.5\%, \notag
%
%
\end{eqnarray}
with corresponding values for the neutron using charge symmetry.
Using other deuteron~\cite{Wiringa:1994wb, Machleidt:2000ge, Gross:2008ps, Gross:2010qm} and $\hel$~\cite{Schulze:1992mb} wave function models does not change our conclusions.

For the parameterization of the off-shell functions,
we take the same form as for the PDFs in Eq.~(\ref{eq.param}), and assume that all quark flavors for the off-shell functions except $\delta u$ and $\delta d$ are zero at the input scale.
The $\delta q_{N/A}$ functions evolve with $Q^2$ in the same way as the on-shell PDFs.
In our fits, we treat $N$, $\alpha$, and $\beta$ as free parameters, and without loss of generality set $\gamma=0$, so that the parameter $\eta$ is fixed by the sum rules.

{\it Quality of fit.}---\
In addition to the new {\footnotesize MARATHON} data, we fit also $F_2$ data from fixed target experiments on $p$ and $D$ from BCDMS~\cite{BCDMS:1989qop}, NMC~\cite{NewMuon:1996fwh, NewMuon:1996uwk}, SLAC~\cite{Whitlow:1991uw}, and Jefferson Lab~\cite{JeffersonLabE00-115:2009jll,Tkachenko:2014byy}, as well as the reduced neutral and charged current proton cross sections from the combined H1/ZEUS analysis from HERA~\cite{H1:2015ubc}, all with kinematic constraints $W^2 > 3.0~{\rm GeV}^2$ and $Q^2 > m_c^2$.
We also include for the first time in a global QCD analysis the Jefferson Lab E03-103 data on $\hel/D$~\cite{Seely:2009gt}.
Drell-Yan di-muon data in $pp$ and $pD$ collisions from the Fermilab E866~\cite{NuSea:1998kqi, NuSea:2001idv} and E906 experiments~\cite{SeaQuest:2021zxb} are included, and for weak vector boson mediated processes we use reconstructed $Z/\gamma^*$ cross-sections and $W^{\pm}$ asymmetries from the Tevatron~\cite{CDF:2010vek, D0:2007djv, CDF:2009cjw, D0:2013lql}, as well as inclusive $W^{\pm}$-lepton asymmetries~\cite{Ringer:2015oaa} from CMS~\cite{CMS:2011bet, CMS:2012ivw, CMS:2013pzl, CMS:2016qqr}, LHCb~\cite{LHCb:2014liz, LHCb:2016nhs}, and STAR~\cite{STAR:2020vuq}.
Also fitted are jet production data from the Tevatron~\cite{D0:2011jpq, CDF:2007bvv} and STAR~\cite{STAR:2006opb}.

\begin{figure}[t]
\includegraphics[width=0.49\textwidth]{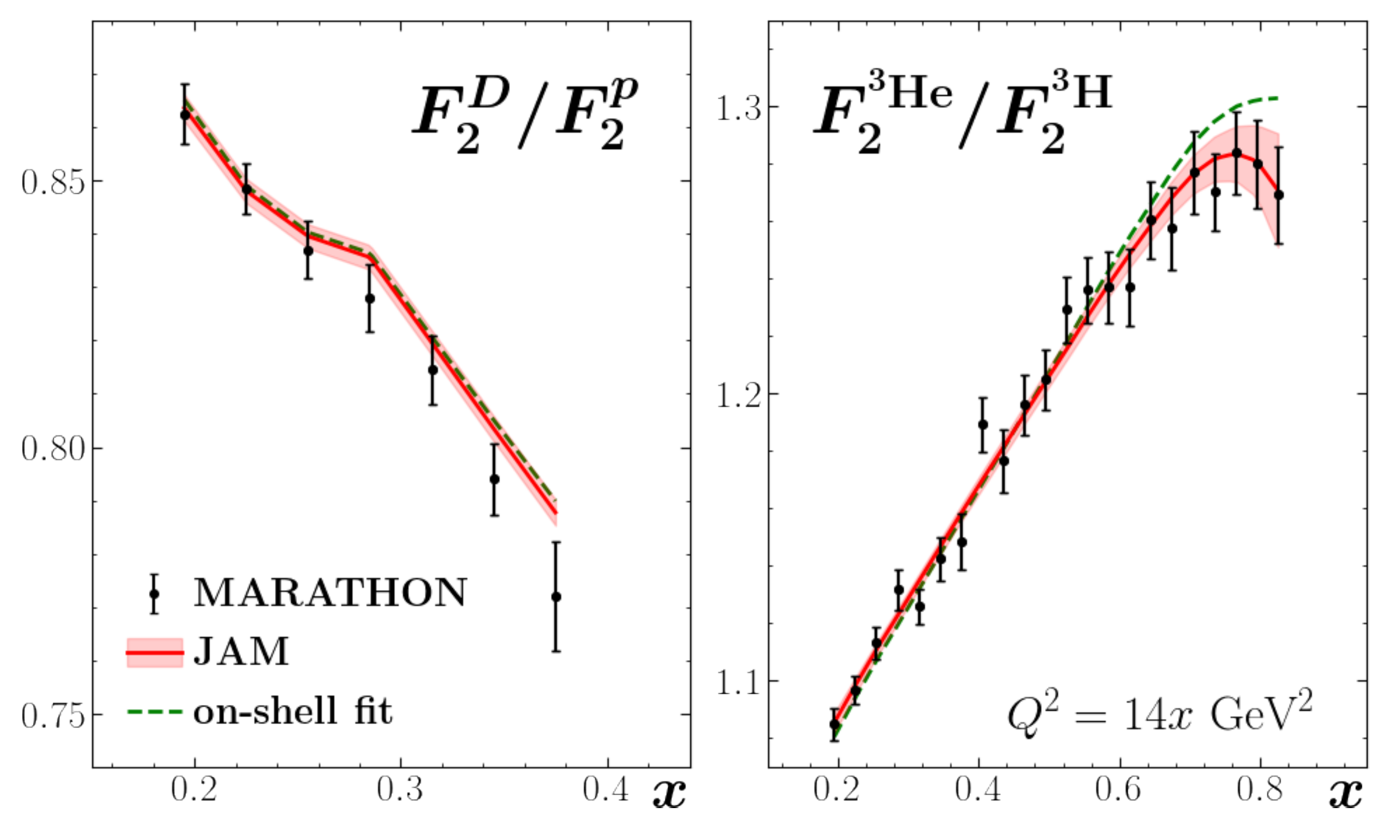}
\caption{Ratios $F_2^D/F_2^p$ (left) and $F_2^{\hel}/F_2^{\tri}$ (right) from \mar\cite{MARATHON:2021vqu} (black circles) at the experimental kinematics $Q^2=14 x$~GeV$^2$ compared with the full JAM fit (red solid lines and $1\sigma$ uncertainty bands) and with an on-shell fit (green dashed lines) which sets the off-shell corrections to zero.}
\label{f.marathon}
\end{figure}

\begin{table}[b]
\caption{Summary of the $\chi^2$ values per number of points $N_{\rm dat}$ for the data used in this analysis. The {\footnotesize MARATHON} and JLab E03-103 $\hel/D$ are separated from the rest of the fixed target data, and their fitted normalizations are shown.}
%
\begin{tabular}{l r c c}
\hhline{====}
process & $N_{\rm dat}$ & ~~\chidof & ~fitted norm.~            \\ \hline
DIS       & & & \\
~~~{\scriptsize MARATHON} {\footnotesize $\hel/\tri$}
                            & 22      & 0.63    & 1.007(6)  \\
~~~{\scriptsize MARATHON} {\footnotesize $D$}/$p$
                            & 7       & 0.95    & 1.019(4)  \\
~~~{\footnotesize JLab E03-103} {\footnotesize $\hel$}/{\footnotesize $D$}
                            & 16      & 0.25    & ~\,1.006(10)  \\
~~~other fixed target       & ~~~2678 & 1.05    &       \\
~~~{\scriptsize HERA}       & 1185    & 1.27    &       \\ 
Drell-Yan                   & 205     & 1.20    &       \\
$W$-lepton asym.            & 70      & 0.81    &       \\
$W$ charge asym.            & 27      & 1.14    &       \\
$Z$ rapidity                & 56      & 1.04    &       \\
jet                         & 200     & 1.11    &       \\ \hline
\textbf{total} & {\bf 4466} & {\bf 1.11}\\
\hhline{====}
\end{tabular} 
\label{t.chi2}
\end{table}

The results of our Monte Carlo analysis are summarized in Table~\ref{t.chi2}.
The overall $\chi^2$ per datum of 1.11 shows that the data are described well.
The resulting fits to the \mar $F_2^D/F_2^p$ and $F_2^{\hel}/F_2^{\tri}$ data are shown in \fref{f.marathon}.
For $D/p$ we are able to fit the data well with a fitted normalization of $1.019(4)$.
For the $\hel/\tri$ ratio, the description of the high-$x$ data improves with the inclusion of off-shell corrections, with the \chidof increasing significantly to 1.29 when the off-shell corrections are switched off.
This dataset displays by far the largest sensitivity to off-shell corrections, and thus is critical for the extraction of these effects.
In Ref.~\cite{MARATHON:2021vqu} a normalization of $1.025(7)$ was included for this dataset based on results from the KP model~\cite{Kulagin:2004ie}, which assumes that $R(D)$ and $\mathcal{R}$ are unity at $x = 0.31$.
To avoid this model bias, we remove this normalization from the data and instead allow the global fit to determine the normalization.
Our fitted value of $1.007(6)$ is in disagreement with the value from the KP model.

\begin{figure}[t]
\includegraphics[width=0.48\textwidth]{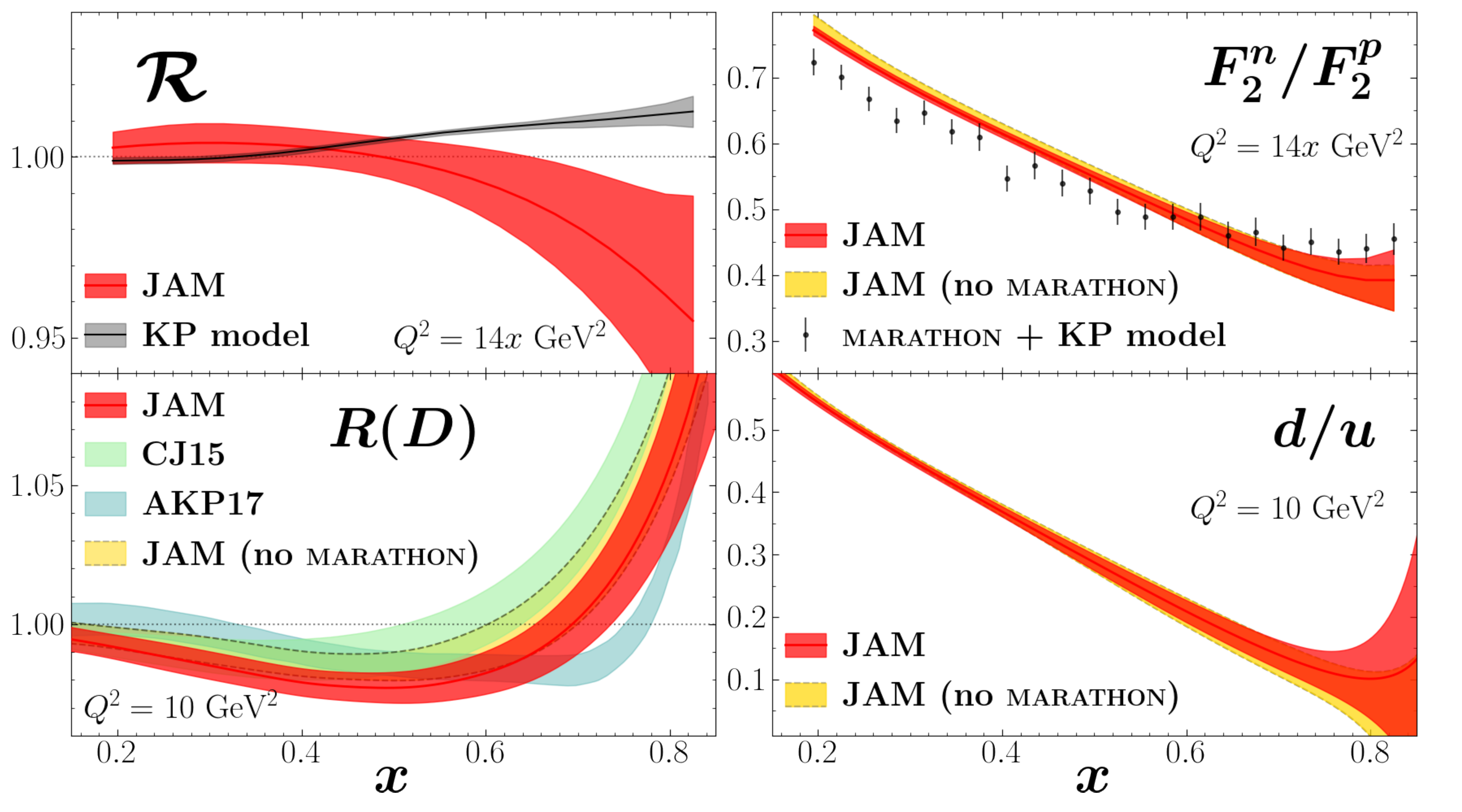}
\caption{Results from the present JAM analysis including \mar data (red bands) for the super-ratio ${\cal R}$ (top left), $F_2^n/F_2^p$ ratio (top right), deuteron EMC ratio $R(D)$ (bottom left), and the $d/u$ ratio (bottom right), compared with those without the \mar data (yellow bands). The super-ratio ${\cal R}$ is compared with the KP model input (gray band) used to extract the $F_2^n/F_2^p$ ratio in \cite{MARATHON:2021vqu}. The deuteron EMC ratio $R(D)$ is also compared with that from CJ15~\cite{Accardi:2016qay} (green band) and AKP17~\cite{Alekhin:2017fpf} (light blue band).}
\label{f.extraction}
\end{figure}

{\it QCD analysis.}---\
The final results of our extraction based on over 1,000 Monte Carlo samples are illustrated in Fig.~\ref{f.extraction}.
For the super-ratio $\mathcal{R}$, our analysis shows that it is consistent with unity until $x \approx 0.7$, at which point it dips and reaches a mean value of 0.96 at $x=0.825$.
The uncertainties on the super-ratio range from $\pm 0.4\%$ at low $x$ up to $\pm 3.5\%$ at the highest $x$.
Without the \mar data the uncertainties on $\mathcal{R}$ (not shown in Fig.~\ref{f.extraction}) vary between 1.3\% and 6.5\%.
This improvement demonstrates that the $\hel/\tri$ data provide a significant amount of information on the super-ratio.
Our results disagree with the KP model~\cite{Kulagin:2004ie}, which predicts a rise to $\mathcal{R} = 1.01$ at $x = 0.825$~\cite{MARATHON:2021vqu}. 
They also suggest that the uncertainties from the KP model, which are an order of magnitude smaller than our extraction even after the inclusion of the \mar data, are significantly underestimated.

Related to the disagreements in the super-ratio, we also find differences between our result for $F_2^n/F_2^p$ and the extraction in Ref.~\cite{MARATHON:2021vqu} based on the KP model.
We find that while the \mar data lowers the central value at low $x$ for the $n/p$ ratio, the central value is still well above the KP model extraction.
At high $x$ values the disagreements are smaller and the inclusion of the \mar data brings our result slightly closer to the KP model extraction.

The impact on the $d/u$  ratio from the inclusion of the \mar data is seen to be very small. 
The small changes for $d/u$ at high $x$ combined with the large differences between the on-shell and off-shell fits at high $x$ (see \fref{f.marathon}) illustrate an important point: Due to the strong constraints placed on the $d/u$ ratio by vector boson production data, and in particular the $W$ asymmetry data from CDF~\cite{CDF:2009cjw} and D0~\cite{D0:2013lql}, the high-$x$ \mar data primarily provide new information on nuclear effects, such as the off-shell corrections, which are most relevant in that region.

For the deuteron EMC ratio $R(D)$, in the intermediate-$x$ region our result is generally in agreement with the CJ15 extraction~\cite{Accardi:2016qay}, while at high $x$ it is between the CJ15 the AKP17~\cite{Alekhin:2017fpf} fits.
Notably, we do not see a strong indication for a unity crossing at $x = 0.31$, as was assumed in Ref.~\cite{MARATHON:2021vqu}.
The inclusion of the \mar $D/p$ data reduces the ratio in the range $0.2 < x < 0.4$.

\begin{figure}[t]
\includegraphics[width=0.48\textwidth]{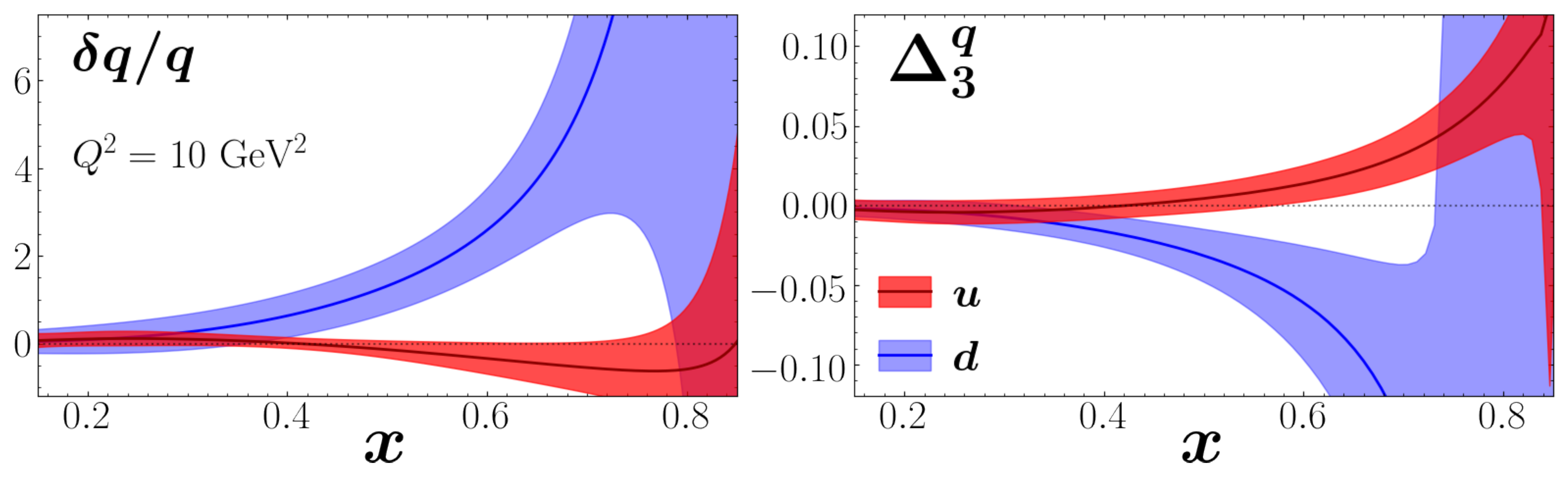}
\caption{Ratio of off-shell to on-shell PDFs $\delta q/q$ (left) and the difference between proton valence quarks in $\hel$ and $\tri$ normalized to the sum, $\Delta_3^q$ (right), for valence $u$ (red bands) and $d$ (blue bands) quarks, at $Q^2 = 10~{\rm GeV}^2$.}
\label{f.isovector}
\end{figure}

The impact of the \mar data on the off-shell corrections $\delta u$ and $\delta d$ is shown in \fref{f.isovector}.
In particular, whereas in the KP model~\cite{Kulagin:2004ie, MARATHON:2021vqu} the proton and neutron off-shell effects are set equal, in our analysis we allow flavor dependence of the effects to be determined from the global fit.
Indeed, we find that while the $\delta u/u$ ratio is consistent with zero, for the $d$ quark the $\delta d/d$ ratio is enhanced at large values of $x$.

An even more direct way of quantifying this effect is to compare the PDFs in the proton bound in $\hel$ and in $\tri$, defining the quantity
\begin{eqnarray}
\Delta_3^q &\equiv& \frac{q_{p/\tri}-q_{p/\hel}}{q_{p/\tri}+q_{p/\hel}},
\label{eq.Delta3q}
\end{eqnarray}
which measures the strength of the isovector EMC effect for $q=u$ and $d$ quarks. 
Since $\hel$ and $\tri$ are mirror nuclei, the ratio $\Delta_3^q$ would vanish if the nuclear corrections were purely isoscalar.
Instead, the behavior in \fref{f.isovector} indicates some deviations from zero for $\Delta_3^d$ at $x \gtrsim 0.4$ and for $\Delta_3^u$ at $x \gtrsim 0.6$.
The fact that the $\Delta_3^q$ are nonzero and of opposite sign for $u$ and $d$ quarks suggests the presence of an isovector component to the EMC effect.
This effect is not taken into account in standard nuclear PDF analyses \cite{Kovarik:2015cma, Eskola:2016oht, NNPDF:2019vjt, Walt:2019slu} which assume $u^{p/A} = d^{n/A}$, and thus may impact not only all nuclear PDF fits, but also numerical calculations that utilize nuclear PDFs in quark-gluon plasma simulations in heavy-ion collisions or neutrino-nucleus interactions in high-energy astrophysics.

{\it Outlook.}---\ 
Our findings are the first indication of an isovector effect in nuclear structure functions, and demonstrate the power of combining the \mar $\hel/\tri$ data with a global QCD analysis to provide simultaneous information on PDFs and nuclear effects in $A \leq 3$ nuclei.
Additional information on the nuclear EMC effects in $^3$He and $^3$H separately will come from $^3$He/$D$ and $^3$H/$D$ ratios measured by {\footnotesize MARATHON}, which are expected to be analyzed in the near future.

Beyond this, constraints on neutron structure, and the $d/u$ PDF ratio at large $x$, will come from the BONuS experiment at Jefferson Lab, which tags spectator protons in semi-inclusive DIS from the deuteron.
Future data on DIS from asymmetric nuclei may also provide further information on the isospin dependence of nuclear effects on structure functions. \\

\begin{acknowledgments}
We thank F.~Ringer and W.~Vogelsang for the code used for calculating the $W$-lepton cross sections, and A.~Accardi, P.~Barry, I.~Fernando and Y.~Zhou for helpful discussions.
This work was supported by the US Department of Energy Contract No.~DE-AC05-06OR23177, under which Jefferson Science Associates, LLC operates Jefferson Lab, and the National Science Foundation under grant number PHY-1516088.
The work of C.C. and A.M. was supported by the US Department of Energy, Office of Science, Office of Nuclear Physics, within the framework of the TMD Topical Collaboration, and by Temple University (C.C.).
The work of A.W.T. was supported by the University of Adelaide and by the Australian Research Council through the Discovery Project DP180100497.
The work of N.S. was supported by the DOE, Office of Science, Office of Nuclear Physics in the Early Career Program.
\end{acknowledgments}



\begin{thebibliography}{99}

\bibitem{Jimenez-Delgado:2013sma}
P.~Jimenez-Delgado, W.~Melnitchouk and J.~F.~Owens,
\href{http://doi.org/10.1088/0954-3899/40/9/093102}
{J. Phys. G \textbf{40}, 093102 (2013)}.

\bibitem{Ethier:2020way}
J.~J.~Ethier and E.~R.~Nocera,
\href{http://doi.org/10.1146/annurev-nucl-011720-042725}
{Ann. Rev. Nucl. Part. Sci. \textbf{70}, 43 (2020)}.

\bibitem{Melnitchouk:1995fc}
W.~Melnitchouk and A.~W.~Thomas,
\href{http://doi.org/10.1016/0370-2693(96)00292-4}
{Phys. Lett. B \textbf{377}, 11 (1996)}.

\bibitem{Kuhlmann:1999sf}
S.~Kuhlmann {\it et al.},
\href{http://doi.org/10.1016/S0370-2693(00)00164-7}
{Phys. Lett. B \textbf{476}, 291 (2000)}.

\bibitem{Accardi:2016qay}
A.~Accardi, L.~T.~Brady, W.~Melnitchouk, J.~F.~Owens, and N.~Sato,
\href{https://doi.org/10.1103/PhysRevD.93.114017}
{Phys. Rev. D {\bf 93}, 114017 (2016)}.

\bibitem{Alekhin:2017fpf}
S.~I.~Alekhin, S.~A.~Kulagin and R.~Petti,
\href{https://doi.org/10.1103/PhysRevD.96.054005}
{Phys. Rev. D \textbf{96}, 054005 (2017)}.

\bibitem{Arrington:2011qt}
J.~Arrington, J.~G.~Rubin and W.~Melnitchouk,
\href{https://doi.org/10.1103/PhysRevLett.108.252001}
{Phys. Rev. Lett. \textbf{108}, 252001 (2012)}.

\bibitem{Baillie:2011za}
N.~Baillie \textit{et al.},
\href{https://doi.org/10.1103/PhysRevLett.108.142001}
{Phys. Rev. Lett. \textbf{108}, 142001 (2012)}.

\bibitem{Tkachenko:2014byy}
S.~Tkachenko \textit{et al.},
\href{https://doi.org/10.1103/PhysRevC.89.045206}
{Phys. Rev. C \textbf{89}, 045206 (2014)}.

\bibitem{Brady:2011hb}
L.~Brady, A.~Accardi, W.~Melnitchouk and J.~F.~Owens,
\href{https://doi.org/10.1007/JHEP06(2012)019}
{JHEP \textbf{06} (2012) 019}.

\bibitem{Afnan:2000uh}
I.~R.~Afnan {\it et al.},
\href{http://doi.org/10.1016/S0370-2693(00)01012-1}
{Phys. Lett. B \textbf{493}, 36 (2000)}.

\bibitem{Afnan:2003vh}
I.~R.~Afnan, F.~R.~P.~Bissey, J.~Gomez, A.~T.~Katramatou, S.~Liuti, W.~Melnitchouk, G.~G.~Petratos and A.~W.~Thomas,
\href{http://doi.org/10.1103/PhysRevC.68.035201}
{Phys. Rev. C \textbf{68}, 035201 (2003)}.

\bibitem{EuropeanMuon:1983wih}
J.~J.~Aubert \textit{et al.},
\href{https://doi.org/10.1016/0370-2693(83)90437-9}
{Phys. Lett. B \textbf{123}, 275-278 (1983)}.

\bibitem{MARATHON:2021vqu}
D.~Abrams {\it et al.},
\href{https://arxiv.org/abs/2104.05850}{arXiv:2104.05850 [hep-ex]}.

\bibitem{Kulagin:2004ie}
S.~A.~Kulagin and R.~Petti,
\href{https://doi.org/10.1016/j.nuclphysa.2005.10.011}
{Nucl. Phys. \textbf{A765}, 126 (2006)}.

\bibitem{Gomez:1993ri}
J.~Gomez \textit{et al.},
\href{https://doi.org/10.1103/PhysRevD.49.4348}
{Phys. Rev. D \textbf{49}, 4348 (1994)}.

\bibitem{Kulagin:2010gd}
S.~A.~Kulagin and R.~Petti,
\href{https://doi.org/10.1103/PhysRevC.82.054614}
{Phys. Rev. C \textbf{82}, 054614 (2010)}.

\bibitem{Sato:2019yez}
N.~Sato, C.~Andres, J.~J.~Ethier, and W.~Melnitchouk,
\href{https://doi.org/10.1103/PhysRevD.101.074020}
{Phys. Rev. D \textbf{101}, 074020 (2020)}.

\bibitem{Moffat:2021dji}
E.~Moffat, W.~Melnitchouk, T.~Rogers and N.~Sato,
\href{https://doi.org/10.1103/PhysRevD.104.016015}
{Phys. Rev. D {\bf 104}, 016015 (2021)}.

\bibitem{Sato:2016tuz}
N.~Sato, W.~Melnitchouk, S.~E.~Kuhn, J.~J.~Ethier and A.~Accardi,
\href{https://doi.org/10.1103/PhysRevD.93.074005}
{Phys. Rev. D {\bf 93}, 074005 (2016)}.

\bibitem{Collins:1989gx}
J.~C.~Collins, D.~E.~Soper and G.~F.~Sterman,
\href{https://doi.org/10.1142/9789814503266\_0001}
{Adv. Ser. Direct. High Energy Phys. \textbf{5}, 1 (1989)}.

\bibitem{Aivazis:1993kh}
M.~A.~G.~Aivazis, F.~I.~Olness and W.-K.~Tung,
\href{https://doi.org/10.1103/PhysRevD.50.3085}
{Phys. Rev. D \textbf{50}, 3085 (1994)}.

\bibitem{Moffat:2019qll}
E.~Moffat, T.~C.~Rogers, W.~Melnitchouk, N.~Sato and F.~Steffens,
\href{https://doi.org/10.1103/PhysRevD.99.096008}
{Phys. Rev. D \textbf{99}, 096008 (2019)}.

\bibitem{Melnitchouk:1993nk}
W.~Melnitchouk, A.~W.~Schreiber and A.~W.~Thomas,
\href{https://doi.org/10.1103/PhysRevD.49.1183}
{Phys. Rev. D \textbf{49}, 1183 (1994)}.

\bibitem{Melnitchouk:1994rv}
W.~Melnitchouk, A.~W.~Schreiber and A.~W.~Thomas,
\href{https://doi.org/10.1016/0370-2693(94)91550-4}
{Phys. Lett. B \textbf{335}, 11 (1994)}.

\bibitem{Kulagin:1994fz}
S.~A.~Kulagin, G.~Piller and W.~Weise,
\href{https://doi.org/10.1103/PhysRevC.50.1154}
{Phys. Rev. C \textbf{50}, 1154 (1994)}.

\bibitem{Kulagin:1994cj}
S.~A.~Kulagin, W.~Melnitchouk, G.~Piller and W.~Weise,
\href{http://doi.org/10.1103/PhysRevC.52.932}
{Phys. Rev. C \textbf{52}, 932 (1995)}.

\bibitem{Ehlers:2014jpa}
P.~J.~Ehlers, A.~Accardi, L.~T.~Brady and W.~Melnitchouk,
\href{http://doi.org/10.1103/PhysRevD.90.014010}
{Phys. Rev. D \textbf{90}, 014010 (2014)}.

\bibitem{Tropiano:2018quk}
A.~J.~Tropiano, J.~J.~Ethier, W.~Melnitchouk and N.~Sato,
\href{http://doi.org/10.1103/PhysRevC.99.035201}
{Phys. Rev. C \textbf{99}, 035201 (2019)}.

\bibitem{Seely:2009gt}
J.~Seely \textit{et al.},
\href{https://doi.org/10.1103/PhysRevLett.103.202301}
{Phys. Rev. Lett. \textbf{103}, 202301 (2009)}.

\bibitem{Londergan:2009kj}
J.~T.~Londergan, J.~C.~Peng and A.~W.~Thomas,
\href{https://doi.org/10.1103/RevModPhys.82.2009}
{Rev. Mod. Phys. \textbf{82}, 2009 (2010)}.

\bibitem{Mineo:2003vc}
H.~Mineo, W.~Bentz, N.~Ishii, A.~W.~Thomas and K.~Yazaki,
\href{https://doi.org/10.1016/j.nuclphysa.2004.02.011}
{Nucl. Phys. \textbf{A735}, 482 (2004)}.

\bibitem{Cloet:2009qs}
I.~C.~Clo\"{e}t, W.~Bentz and A.~W.~Thomas,
\href{https://doi.org/10.1103/PhysRevLett.102.252301}
{Phys. Rev. Lett. \textbf{102}, 252301 (2009)}.

\bibitem{Lacombe:1981eg}
M.~Lacombe, B.~Loiseau, R.~Vinh Mau, J.~Cote, P.~Pires and R.~de Tourreil,
\href{https://doi.org/10.1016/0370-2693(81)90659-6}
{Phys. Lett. B \textbf{101}, 139 (1981)}.

\bibitem{Kievsky:1996gz}
A.~Kievsky, E.~Pace, G.~Salme and M.~Viviani,
\href{https://doi.org/10.1103/PhysRevC.56.64}
{Phys. Rev. C \textbf{56}, 64 (1997)}.

\bibitem{Wiringa:1994wb}
R.~B.~Wiringa, V.~G.~J.~Stoks and R.~Schiavilla,
\href{https://doi.org/10.1103/PhysRevC.51.38}
{Phys. Rev. C \textbf{51}, 38 (1995)}.

\bibitem{Machleidt:2000ge}
R.~Machleidt,
\href{https://doi.org/10.1103/PhysRevC.63.024001}
{Phys. Rev. C \textbf{63}, 024001 (2001)}.

\bibitem{Gross:2008ps}
F.~Gross and A.~Stadler,
\href{https://doi.org/10.1103/PhysRevC.78.014005}
{Phys. Rev. C \textbf{78}, 014005 (2008)}.

\bibitem{Gross:2010qm}
F.~Gross and A.~Stadler,
\href{https://doi.org/10.1103/PhysRevC.82.034004}
{Phys. Rev. C \textbf{82}, 034004 (2010)}.

\bibitem{Schulze:1992mb}
R.~W.~Schulze and P.~U.~Sauer,
\href{https://doi.org/10.1103/PhysRevC.48.38}
{Phys. Rev. C \textbf{48}, 38 (1993)}.

\bibitem{BCDMS:1989qop}
A. C. Benvenuti {\it et al.},
\href{https://doi.org/10.1016/0370-2693(89)91637-7}
{Phys. Lett. B {\bf 223}, 485 (1989)}; 
{\it ibid.} \href{https://doi.org/10.1016/0370-2693(90)91231-Y}
{B {\bf 237}, 592 (1990)}.

\bibitem{NewMuon:1996fwh}
M. Arneodo {\it et al.},
\href{https://doi.org/10.1016/S0550-3213(96)00538-X}
{Nucl. Phys. {\bf B483}, 3 (1997)}.

\bibitem{NewMuon:1996uwk}
M. Arneodo {\it et al.},
\href{https://doi.org/10.1016/S0550-3213(96)00673-6}
{Nucl. Phys. {\bf B487}, 3 (1997)}.

\bibitem{Whitlow:1991uw}
L. W. Whitlow {\it et al.},
\href{https://doi.org/10.1016/0370-2693(92)90672-Q}
{Phys. Lett. B {\bf 282}, 475 (1992)}.

\bibitem{JeffersonLabE00-115:2009jll}
S. P. Malace {\it et al.},
\href{https://doi.org/10.1103/PhysRevC.80.035207}
{Phys. Rev. C {\bf 80}, 035207 (2009)}.

\bibitem{H1:2015ubc}
H. Abramowicz {\it et al.},
\href{https://doi.org/10.1140/epjc/s10052-015-3710-4}
{Eur. Phys. J. C {\bf 75}, 580 (2015)}.

\bibitem{NuSea:1998kqi}
E. A. Hawker {\it et al.},
\href{https://doi.org/10.1103/PhysRevLett.80.3715}
{Phys. Rev. Lett. {\bf 80}, 3715 (1998)};
J. Webb, Ph.D. Thesis, New Mexico State University (2002),
\href{https://arxiv.org/abs/hep-ex/0301031}
{arXiv:hep-ex/0301031}.

\bibitem{NuSea:2001idv}
R.~S.~Towell \textit{et al.},
\href{https://doi.org/10.1103/PhysRevD.64.052002}
{Phys. Rev. D \textbf{64}, 052002 (2001)}.

\bibitem{SeaQuest:2021zxb}
J.~Dove \textit{et al.},
\href{https://doi.org/10.1038/s41586-021-03282-z}
{Nature \textbf{590}, 561 (2021)}.

\bibitem{CDF:2010vek}
T. Aaltonen {\it et al.},
\href{https://doi.org/10.1016/j.physletb.2010.06.043}
{Phys. Lett. B {\bf 692}, 232 (2010)}.

\bibitem{D0:2007djv}
V. M. Abazov {\it et al.},
\href{https://doi.org/10.1103/PhysRevD.76.012003}
{Phys. Rev. D {\bf 76}, 012003 (2007)}.

\bibitem{CDF:2009cjw}
T. Aaltonen {\it et al.},
\href{https://doi.org/10.1103/PhysRevLett.102.181801}
{Phys. Rev. Lett. {\bf 102}, 181801 (2009)}.

\bibitem{D0:2013lql}
V. M. Abazov {\it et al.},
\href{https://doi.org/10.1103/PhysRevLett.112.151803}
{Phys. Rev. Lett. {\bf 112}, 151803 (2014)}; 
{\it ibid.} 
\href{https://doi.org/10.1103/PhysRevLett.114.049901}
{{\bf 114}, 049901 (2015)}.

\bibitem{Ringer:2015oaa}
F. Ringer, W. Vogelsang,
\href{https://doi.org/10.1103/PhysRevD.91.094033}
{Phys. Rev. D {\bf 91}, 094033 (2015)}.

\bibitem{CMS:2011bet}
S. Chatrchyan \textit{et al.},
\href{ https://doi.org/10.1007/JHEP04(2011)050}
{JHEP {\bf 04} (2011) 050}.

\bibitem{CMS:2012ivw}
S. Chatrchyan \textit{et al.},
\href{https://doi.org/10.1103/PhysRevLett.109.111806}
{Phys. Rev. Lett. {\bf 109}, 111806 (2012)}.

\bibitem{CMS:2013pzl}
S. Chatrchyan \textit{et al.},
\href{https://doi.org/10.1103/PhysRevD.90.032004}
{Phys. Rev. D {\bf 90}, 032004 (2014)}.

\bibitem{CMS:2016qqr}
V. Khachatryan \textit{et al.},
\href{https://doi.org/10.1140/epjc/s10052-016-4293-4}
{Eur. Phys. J. C {\bf 76}, 469 (2016)}.

\bibitem{LHCb:2014liz}
R.~Aaij \textit{et al.}, 
\href{https://doi.org/10.1007/JHEP12(2014)079}
{JHEP \textbf{12} (2014) 079}.

\bibitem{LHCb:2016nhs}
R.~Aaij \textit{et al.}, 
\href{https://doi.org/10.1007/JHEP01(2016)155}
{JHEP \textbf{01} (2016) 155}.

\bibitem{STAR:2020vuq}
J.~Adam \textit{et al.}, 
\href{https://doi.org/10.1103/PhysRevD.103.012001}
{Phys. Rev. D \textbf{103}, 012001 (2021)}.

\bibitem{D0:2011jpq}
V. M. Abazov {\it et al.},
\href{https://doi.org/10.1103/PhysRevLett.101.062001}
{Phys. Rev. Lett. {\bf 101}, 062001 (2008)}.

\bibitem{CDF:2007bvv}
A. Abulencia {\it et al.},
\href{https://doi.org/10.1103/PhysRevD.75.092006}
{Phys. Rev. D {\bf 75}, 092006 (2007)}
[Erratum: 
\href{https://doi.org/10.1103/PhysRevD.75.119901}
{Phys. Rev. D {\bf 75}, 119901 (2007)}].

\bibitem{STAR:2006opb}
B. I. Abelev {\it et al.},
\href{https://doi.org/10.1103/PhysRevLett.97.252001}
{Phys. Rev. Lett. {\bf 97}, 252001 (2006)}.

\bibitem{Kovarik:2015cma}
K.~Kovarik {\it et al.}, 
\href{https://doi.org/10.1103/PhysRevD.93.085037}
{Phys. Rev. D \textbf{93}, 085037 (2016)}.

\bibitem{Eskola:2016oht}
K.~J.~Eskola, P.~Paakkinen, H.~Paukkunen and C.~A.~Salgado,
\href{https://doi.org/10.1140/epjc/s10052-017-4725-9}
{Eur. Phys. J. C \textbf{77}, 163 (2017)}.

\bibitem{NNPDF:2019vjt}
R.~Abdul Khalek \textit{et al.}, 
\href{https://doi.org/10.1140/epjc/s10052-019-7364-5}
{Eur. Phys. J. C {\bf 79}, 838 (2019)}.

\bibitem{Walt:2019slu}
M.~Walt, I.~Helenius and W.~Vogelsang,
\href{https://doi.org/10.1103/PhysRevD.100.096015}
{Phys. Rev. D \textbf{100}, 096015 (2019)}.


\end{thebibliography}
\end{document}